\documentclass[aps,prc,preprint,showpacs,floatfix]{revtex4-1}
\usepackage[dvips]{color}
\usepackage{graphicx}
\usepackage{bm}

\begin{document}

\title{Alpha decay of nuclei interesting for synthesis of $Z=119, 120$
isotopes}

\author{D. N. Poenaru$^{1,2}$, and R.  A.  Gherghescu$^{1,2}$}

\email[]{poenaru@fias.uni-frankfurt.de, radu@theory.nipne.ro}

\affiliation{$^1$Horia Hulubei National Institute of Physics and Nuclear
Engineering (IFIN-HH), \\P.O. Box MG-6, RO-077125 Bucharest-Magurele,
Romania, and $^2$Frankfurt Institute for Advanced Studies, Johann Wolfgang
Goethe University, Ruth-Moufang-Str. 1, D-60438 Frankfurt, Germany}

\date{ }

\begin{abstract}

Four groups (even-even, even-odd, odd-even and odd-odd) of heavy and
super-heavy nuclei are of interest for the synthesis of the isotopes with
$Z=119, 120$.  We report calculations of $\alpha$~decay half-lives using
four models: AKRA (Akrawy); ASAF (Analytical Super-Asymmetric Fission); UNIV
(Universal Formula), and semFIS (Semi-empirical formula based on Fission
Theory).  We compare the experimental $Q_\alpha$ values either with AME16
atomic mass evaluation (whenever available) and with the theoretical model
WS4, able to give masses of not yet measured nuclides.  For $^{92,94}$Sr
cluster radioactivity of $^{300,302}$120 we predict a branching ratio
relative to $\alpha$~decay of -0.10 and 0.49, respectively, meaning that it
is worth trying to detect such kind of decay modes in competition with
$\alpha$~decay.

\end{abstract}


\maketitle

\section{Introduction}
\label{sec:1}

Super-heavy (SH) nuclei \cite{ham13arnps,khu14prl} with atomic number Z up
to 118, have been produced by two kinds of fusion reactions: (1) almost cold
fusion (with one evaporated neutron) at GSI Germany \cite{hof00rmp,hof11ra}
and RIKEN Japan \cite{mor07jpsjb} based on the doubly magic target
$^{208}$Pb or its neighbour $^{209}$Bi, and (2) hot fusion (with three or
four evaporated neutrons) at JINR Dubna Russia and Livermore Nat. Lab. USA
\cite{oga10prl,oga15rpp} with the, quite expensive, $^{48}$Ca projectile.

From the attempts to synthesize $Z=119, 120$ isotopes
\cite{hof16epja,hof17dub,hof17sanibel} we selected 21 even-even (e-e), 21
even-odd (e-o), 4 odd-even (o-e), and 5 odd-odd (o-o) $\alpha$~emitters. 
Some of them are reported from experiments with different values of
$Q_\alpha$ or half-life $T_{1/2}$.  Consequently the total number of cases
is larger: 29 e-e, 58 e-o, 5 o-e, and 9 o-o.  To these we also added few (3
e-e, 10 o-e, and 4 o-o) other nuclides, members of the $\alpha$~decay chains
of $^{300,302}120$, $^{299,301}120$, $^{297}119$ and $^{300}119$, namely
$^{273}Bh_{107}$, $^{277}Mt_{109}$, $^{281,283,284}Rg_{111}$, 
$^{285,287,288}Nh_{113}$, $^{289,291,292}Mc_{115}$, $^{294}Bh_{116}$,
$^{293,295,296}Ts_{117}$, $^{296,298}Og_{118}$.
We express the half-lives in decimal logarithm of the values
in seconds, $T=\log_{10}T_{1/2}(s)$.
Whenever possible we rely on the latest (AME16) atomic mass evaluation
\cite{men17cpc} 
in order to calculate $Q_\alpha$.  The following nuclides
are not available on AME16 evaluation: {\bf e-e} $^{278}Hs_{108}$;
$^{282}Ds_{110}$; $^{284}Fl_{114}$; $^{290}Fl_{114}$; $^{294}Fl_{114}$;
$^{300}120$ and $^{302}120$; {\bf e-o} $^{277}Hs_{108}$; $^{299}120$ and
$^{301}120$; {\bf o-e} $^{297}119$ and {\bf o-o} $^{278}Bh_{107}$;
$^{282}Mt_{109}$; $^{286}Rg_{111}$; $^{290}Nh_{113}$ and $^{300}119$.
In this way, there is no o-o nuclide remaining on the AME16 table.
For nuclides not available in this table we use the model W4 \cite{wan14plb,
wan10prc}, which was found \cite{wan15prc} to be the best among 20 models. 
Also, in the same Ref., it is mentioned that {\em ``SemFIS2 formula is the
best one to predict the alpha-decay half-lives ...  In addition, the UNIV2
formula with fewest parameters and ...  work well in prediction on the SHN
alpha-decay half-lives''.} We shall use semFIS, UNIV, ASAF
\cite{p83jpl80,p103jp84,p140b89,p172b96,p195b96,p193b97,p302bb10} and AKRA
\cite{akr17ep}.  A computer program \cite{p98cpc82} gives us the possibility
to improve the parameters of the ASAF model in agreement with a given set of
experimental data.  The UNIV (universal curve) model was updated in 2011
\cite{p308prc11}. Nevertheless, for $^{297,299}$119 nuclei we couldn't get
the Q-values by using the model W4, hence in these particular cases the
model KTUY05 \cite{kou05ptp} have been used.
Interesting developments concerning alpha decay, cluster radioactiovity,
spontaneous fission and proton radioactiovity have been recently made 
\cite{zde13prc,pom15ps,zde16epja}.

In the decay modes we are studying, a parent nucleus,$^A$Z , disintegrates
with emission of a light particle, $^{Ae}Z_e$, and a heavy daughter
$^{Ad}Z_d$
\begin{equation}
^AZ \rightarrow ^{Ae}Z_e + ^{Ad}Z_d
\end{equation}
The kinetic energy of the $\alpha$ particle is related to Q-value by the
relationship
\begin{equation}
E_k=QA_d/A
\end{equation}
and Q-value is calculated from the atomic masses using the Einstein's
relationship
\begin{equation}
Q=[M_p - (M_e + M_d)]mc^2 
\end{equation}
where c is the speed of light.

ASAF, fragmentation theory developed by the Frankfurt School, and other
models have been used to predict cluster radioactivities \cite{enc95}.  For
some isotopes of SHs, with $Z>121$, there is a good chance for cluster decay
modes to compete \cite{p315prc12,p309prl11}.

In the following section, models, we shall give some informations concerning
the AKRA, ASAF, UNIV, and semFIS models.  Then in section, released energy,
we shall compare the experimental values of $Q_\alpha$ with those obtained
from AME16 and W4.  In the section, results, we shall compare the half-lives
obtained with the four models, with experimental data.  In conclusion and
outlook we evaluate how useful is any of the four models, and what to do in
order to improve the present aituation.

\section{Models}

The half-life of a parent nucleus $AZ$ against the split into a cluster $A_e
Z_e$ and a daughter $A_d Z_d$
\begin{equation}
T = [(h \ln 2)/(2E_{v})] exp(K_{ov} + K_{s})
\end{equation}
is calculated by using the WKB quasiclassical approximation, according to
which the action integral is expressed as
\begin{equation}
K =\frac{2}{\hbar}\int_{R_a}^{R_b}\sqrt{2B(R)[E(R)-Q]} dR
\label{wkb}
\end{equation}
with $B=\mu$ --- the reduced mass, $K=K_{ov}+K_s$ (overlapping and separated
fragments), and $E(R)$ is the total
deformation energy. $R_a, R_b$ are the turning points, defined by
\begin{equation}
E(R_a)-Q=E(R_b)-Q=0
\end{equation}

\subsection{AKRA}

In Ref.~\cite{ap358ar17}  a new formula was introduced, derived by adding
few parameters to the one developed by G.  Royer \cite{roy10np}.  Three
experimental data sets have been used: A (130 e-e, 119 e-o, 109 o-e, and 96
o-o), set B (188 e-e, 147 e-o, 131 o-e, and 114 o-o), and set C with 136
e-e, 84 e-o, 76 o-e, and 48 o-o alpha emitters.set C with 136 e-e, 84 e-o,
76 o-e, and 48 o-o alpha emitters.  The set A was developed by one of us
(DA), the set B belongs to DNP's group, and the set C was taken from G. 
Royer \cite{roy10np}; few Q-values have been updated using the AME16
evaluation of experimental atomic masses \cite{aud17cpc}.  Comparison with
ASAF, UNIV, and semFIS will be made using both A, B and C data sets.

The Royer formula \cite{roy10np} is defined as
\begin{equation}
T_{1/2} = a + bA^{1/6}\sqrt{Z} + \frac{cZ}{\sqrt{Q_\alpha}}
\end{equation}
with initial parameters $a=-27.657; -28.408; -27.408, and -24.763$,
$b=-0.966; -0.920;$ $-1.038, and -0.907$, and $c=1.522; 1.519; 1.581, and
1.410$ for e-e, e-o, o-e, and o-o, respectively. The rms standard deviation
for 130 e-e, 119 e-o, 109 o-e, and 96 o-o was $\sigma = 0.560, 1.050,
0.871, $ and $0.926$, respectively.

The new relationship is obtained by introducing $I=(N-Z)/A$ and the
new parameters d and e:
\begin{equation}
T_{1/2} = a + bA^{1/6}\sqrt{Z} + \frac{cZ}{\sqrt{Q_\alpha}} + dI + eI^2
\end{equation}
where for the comprehensive set B the parameters a, b, c, d, e are given in
Table~\ref{tabcb}.

Before optimization, with our set of 580 $\alpha$~emitters, and the initial
values of the parameters $a=-27.989, b=-0.940, c=1.532, d=-5.747, e=11.336 $
for even-even nuclei we got the following values of rms standard deviations,
$\sigma = 0.5547$.  After optimization for e-e emitters, with
$a=-27.837,b=-0.94199975, c=1.5343, d=-5.7004, e=8.785$ the agreement was
improved: $\sigma = 0.540$.  

By ruling with the optimized set of 21 e-e, 21 e-o, 13 o-e and 9 o-o
$\alpha$~emitters, we adjusted only the value of parameter e, in order to
get the best fit.  The parameters have been $a=-27.949, -28.215,-26.594,
-23.936$, $b=-0.94199975,-0.861,-1.107,-0.891$,
$c=1.5343,1.53774,1.557,1.404$, $d=-5.7004,-21.145,15.149,-12.420$ and
$e=22.560, -18.200,$ $-77.700, 33.000$. 
We got for even-even nuclei the rms standard 
deviation, $\sigma_{ee} = 2.9892$, for even-odd $\sigma_{eo} = 8.0620$,  for
odd-even $\sigma_{oe} = 3.8810$, and for odd-odd $\sigma_{oo} = 1.366$.

\subsection{ASAF}

We replace in eq.~\ref{wkb} $E(R)-Q$ by $[E(R)-E_{corr}] - Q$.  $E_{corr}$
is a correction energy similar to the Strutinsky shell correction, also
taking into account the fact that Myers-Swiatecki's liquid drop model (LDM)
overestimates fission barrier heights, and the effective inertia in the
overlapping region is different from the reduced mass.  The turning points
of the WKB integral are:
\begin{equation}
R_a = R_i + (R_t - R_i)[(E_v + E^*)/E_b^0]^{1/2}
\end{equation}
\begin{equation}
R_b = R_tE_c \{ 1/2+[1/4+(Q+E_v+E^*)E_l/E_c^2]^{1/2} \} /(Q+E_v+E^*)
\end{equation}
where $E^*$ is the excitation energy concentrated in the separation degree
of freedom, $R_i=R_0-R_e$ is the initial separation distance, $R_t=R_e+R_d$
is the touching point separation distance, $R_j=r_0A_j^{1/3}$ $(j=0, e, d ;
\; r_0=1.2249$~fm) are the radii of parent, emitted and daughter nuclei, and
$E_b^0=E_i-Q$ is the barrier height before correction.  The interaction
energy at the top of the barrier, in the presence of a non negligible angular
momentum, $l\hbar$, is given by:
\begin{equation}
E_i=E_c+E_l=e^2Z_eZ_d/R_t+\hbar^2l(l+1)/(2\mu R_t^2)
\end{equation}
The two terms of the action integral $K$, corresponding to the overlapping
($K_{ov}$) and separated ($K_s$) fragments, are calculated by analytical  
formulas (approximated for $K_{ov}$ and exact for $K_s$ in case of separated
spherical shapes within the LDM):
\begin{equation}
K_{ov}=0.2196(E^0_bA_eA_d/A)^{1/2}(R_t-R_i)\left [  \sqrt{1-b^2}-
b^2 \ln \frac{1+\sqrt{1-b^2}}{b} \right ]
\end{equation}                           
\begin{equation}                         
K_{s}=0.4392[(Q+E_v+E^*)A_eA_d/A]^{1/2}R_b J_{rc} \;   ;  \;
b^2=(E_v+E^*)/E_b^0
\end{equation} 
\begin{eqnarray}
J_{rc}  & = & (c) \arccos \sqrt{(1-c+r)/(2-c)} - [(1-r)(1-c+r)]^{1/2}
                \nonumber \\
        & + & \sqrt{1-c} \ln \left [
  \frac{2\sqrt{(1-c)(1-r)(1-c+r)} +2-2c+cr}{r(2-c)} \right ]
\end{eqnarray}
where $r=R_t/R_b$ and $c=rE_c/(Q+E_v+E^*)$. In the absence of the
centrifugal contribution ($l=0$), one has $c=1$.

The choice $E_{v} = E_{corr}$ allows to get a smaller number of parameters. 
Owing to the exponential dependence, any small variation
of $E_{corr}$ induces a large change of T, and thus plays a more important
role compared to the preexponential factor variation due to $E_{v}$.  Shell
and pairing effects are included in $E_{corr}=a_i(A_e)Q$ ($i=1,2,3,4$ for
even-even, odd-even, even-odd, and odd-odd parent nuclei).  For a given
cluster radioactivity there are four values of the coefficients $a_i$, the
largest for even-even parent and the smallest for the odd-odd one (see
figure~1 of \cite{p160adnd91}).  The shell effects for every cluster
radioactivity is implicitly contained in the correction energy due to its
proportionality with the Q value.
Since 1984, the ASAF model results have been
used to guide the experiments and to stimulate other theoretical works.

In the present case we obtained the following rms standard deviations:
$\sigma_{ee} = 3.397$, for even-odd $\sigma_{eo} = 8.458$,  for
odd-even $\sigma_{oe} = 4.056$, and for odd-odd $\sigma_{oo} = 1.663$

\subsection{UNIV (Universal Formula)}

In cluster radioactivity and $\alpha$-decay the (measurable) decay constant
$\lambda = \ln 2 /T$, can be expressed as a product of three (model
dependent) quantities
\begin{equation}   
\lambda = \nu S P_s
\end{equation}
where $\nu $ is the frequency of assaults on the barrier per second, $S$ is
the preformation probability of the cluster at the nuclear surface, and
$P_s$ is the quantum penetrability of the external potential barrier.  The
frequency $\nu$ remains practically constant, the preformation differs from
one decay mode to another but it is not changed very much for a given
radioactivity, while the general trend of penetrability follows closely that
of the half-life.  

The preformation probability can be calculated within a fission model as a
penetrability of the internal part of the barrier, which corresponds to
still overlapping fragments.  One may assume as a first approximation, that
preformation probability only depends on the mass number of the emitted
cluster, $S=S(A_e)$.  The next assumption is that $\nu(A_{e}, Z_{e}, A_{d},
Z_{d}) = $~constant.  In this way it was obtained a single straight line
{\em universal curve} on a double logarithmic scale 
\begin{equation} \log T=-\log P_s - 22.169 + 0.598(A_{e}-1) 
\end{equation} 
where 
\begin{equation}
-\log P_s = c_{AZ} \left [\arccos \sqrt{r} - \sqrt{r(1-r)}\right ]
\end{equation}
with $c_{AZ}= 0.22873({\mu}_{A}Z_{d}Z_{e}R_{b})^{1/2}$,
$r=R_{t}/R_{b}$, $R_{t}=1.2249(A_{d}^{1/3}+A_{e}^{1/3})$,
$R_{b}=1.43998Z_{d}Z_{e}/Q$, and $\mu_A = A_d A_e / A$.

Sometimes this universal curve is misinterpreted as being a Geiger-Nuttal
plot.  Nowadays by Geiger-Nuttal diagram one understands a plot of $\log T$
versus $ZQ^{-1/2}$, or versus $Q^{-1/2}$.  
For $\alpha$-decay of even-even nuclei, $A_e=4$, one has
\begin{equation}
\log T = -\log P_s + c_{ee}
\end{equation}
where $c_{ee}=\log S_{\alpha} - \log \nu + \log (\ln 2) = -20.375$.  We can
find new values for $c_{ee}$ and we also can extend the relationship to
even-odd, odd-even, and odd-odd nuclei, by fitting a given set of
experimentally determined alpha decay data.

By adjusting every time the additive constant $c_{ee}$ we obtained the
following rms standard deviations: $\sigma_{ee} = 2.952$ when
$c_{ee}=1.420$, $\sigma_{eo} = 8.146$ when $c_{eo}=-1.600$, $\sigma_{oe} =
3.989$ when $c_{oe}=-0.700$, and $\sigma_{oo} = 1.548$ when $c_{oe}=1.392$.

\subsection{semFIS (Semiempirical relationship based on fission theory of
$\alpha$-decay)}

Mainly the $Z$ dependence was stressed by all formulae, in spite of
strong influence of the neutron shell effects. The neighborhood of
the magic numbers of nucleons is badly described by all these
relationships.
The SemFIS formula  based on the fission theory of $\alpha$-decay gives
\begin{equation}
\log T = 0.43429K_s\chi - 20.446
\label{eq:pf}
\end{equation}
where 
\begin{eqnarray}
K_s & = & 2.52956 Z_{da}[A_{da}/(AQ_{\alpha})]^{1/2}[\arccos \sqrt{x} 
- \sqrt{x(1-x)}] \; ; \nonumber \\
   & x & =0.423Q_{\alpha}(1.5874+A_{da}^{1/3})/Z_{da} 
\end{eqnarray}
and the numerical coefficient $\chi$, close to unity, is a 
second-order polynomial
\begin{equation}
\chi=B_1+B_2y+B_3z+B_4y^2+B_5yz+B_6z^2 
\end{equation}
in the reduced variables $y$ and $z$, expressing the distance from the
closest magic-plus-one neutron and proton numbers $N_i$ and $Z_i$:
\begin{equation}
y \equiv (N-N_i)/(N_{i+1} - N_i) \; ; \; N_i < N \le N_{i+1}
\end{equation}
\begin{equation}
z \equiv (Z-Z_i)/(Z_{i+1} - Z_i) \; ; \; Z_i < Z \le Z_{i+1}
\label{eq:pl}
\end{equation}
with $N_i=...., 51, 83, 127, 185, 229, .....$ , $Z_i=...., 29, 51, 83, 115,
.....$ , and $Z_{da}=Z-2$ , $A_{da}=A-4$ .  The coefficients $B_i$ are obtained
by fit with experimental data, using a computer program making automatically
the best fit \cite{p98cpc82}.  Better agreement with experimental results
are obtained in the region of superheavy nuclei by introducing other values
of the magic numbers plus one unit for protons (suggesting that the next
magic number of protons could be 126 instead of 114): $Z_i=...., 83, 127,
165, .....$

Practically for even-even nuclei, the increased errors in the neighborhood
of $N=126$, present in all other cases, are smoothed out by SemFIS~formula
using the second order polynomial approximation for $\chi$.  They are still
present for the strongest $\alpha$-decays of some even-odd and odd-odd
parent nuclides.  In fact for non-even number of nucleons the structure
effects became very important, and they should be carefully taken into
account for every nucleus, not only globally.  An overall estimation of the
accuracy, gives the standard rms deviation of $\log T$ values:
\begin{equation}
\sigma = \left\{\sum_{i=1}^n[\log
(T_i/T_{exp})]^2/(n-1)\right\}^{1/2}
\end{equation}

The partial $\alpha$-decay half-lives plotted in this figure are lying in
the range of $10^{-7}$ to $10^{25}$ seconds.  One can see the effect of the
spherical and deformed neutron magic numbers of the daughter nuclei
$N_d=126, 152, 162$ particularly clear for even-even and even-odd nuclides. 
For the large set of alpha emitters the following values of the rms errors
have been obtained: $\log T$: 0.19 for SemFIS~formula; 0.33 for the
universal curve; 0.39 for ASAF model, and 0.43 for numerical superasymmetric
(NuSAF) model.

There are many parameters of the SemFIS formula introduced in order to
reproduce the experimental behaviour around the magic numbers of protons and
neutrons, which could be a drawback in the region of light and intermediate
alpha emitters.  

We succeeded to obtain $\sigma_{ee} = 3.178$ when $B1= 0.993119,
B2=-0.0046700, B3= 0.017009, B4= 0.045030, B5= 0.018101, B6=-0.025097$,
$\sigma_{eo}= 9.453$ when $B1= 1.017560, B2=-0.113054, B3= 0.019057, B4=
0.147320, B5= 0.230300, B6=-0.101528$, $\sigma_{oe}= 4.439$ when $B1=
1.000560, B2= 0.010783, B3= 0.050671, B4= 0.013918, B5= 0.043657,
B6=-0.079999$, and $\sigma_{oo}= 2.885$ when $B1= 1.004470, B2=-0.160056,
B3= 0.264857, B4= 0.212332, B5= 0.292664, B6=-0.401158$.

\section{Released Energy}

\subsection{AME16 and W4}

The results obtained with AME16 are showing good agreement with experimental
data, particularly for e-e emitters, folowed by o-e ones.

The largest error (over $|0.1|$) is obtained for: $^{253}Es_{99}$,
$^{269,271}Sg_{106}$, $^{273,275}Hs_{108}$, $^{277,279,281}Ds_{110}$,
$^{283,285}Cn_{112}$, $^{287}Nh_{113}$, $^{285,286,287,288,289}Fl_{114}$,
$^{290,291,292,293}Lv_{116}$, and $^{294,295}Og_{118}$.  The best result
(under $|0.01|$) is obtained for $^{281,284,281,285}Cn_{112}$,
$^{287,288,289}Fl_{114}$, $^{290}Lv_{116}$, $^{294}Og_{118}$, and the almost
perfect $^{291}Lv_{116}$.  Some of the nuclides appear both on the wrong and
the best side, because the input data may be different for the same emitter.

From the results obtained with W4 atomic masses one can see the worst
results (dq over 1 unit) for $^{253}Es_{99}$, $^{265,267}Rf_{104}$,
$^{269,271,273}Sg_{106}$, $^{278}Bh_{107}$,
$^{273,275,277,277,278}Hs_{108}$, $^{282}Mt_{109}$,
$^{277,279,281,282}Ds_{110}$, $^{286}Rg_{111}$, $^{281,283,285}Cn_{112}$,
$^{287}Nh_{113}$, $^{284,285,286,287,288,289,290,291,293,294}Fl_{114}$,
$^{290,291,292,293}Lv_{116}$, and $^{294,295}Og_{118}$. 
The best result is obtained for $^{281,282,285}Cn_{112}$,
$^{286,287,288}Fl_{114}$, $^{290,292}Lv_{116}$, $^{291}Lv_{116}
(dq=-0.954E-06 !!!) $, and $^{294}Og_{118}$.

\begin{figure}[ht]
\begin{center}
\includegraphics[width=15cm]{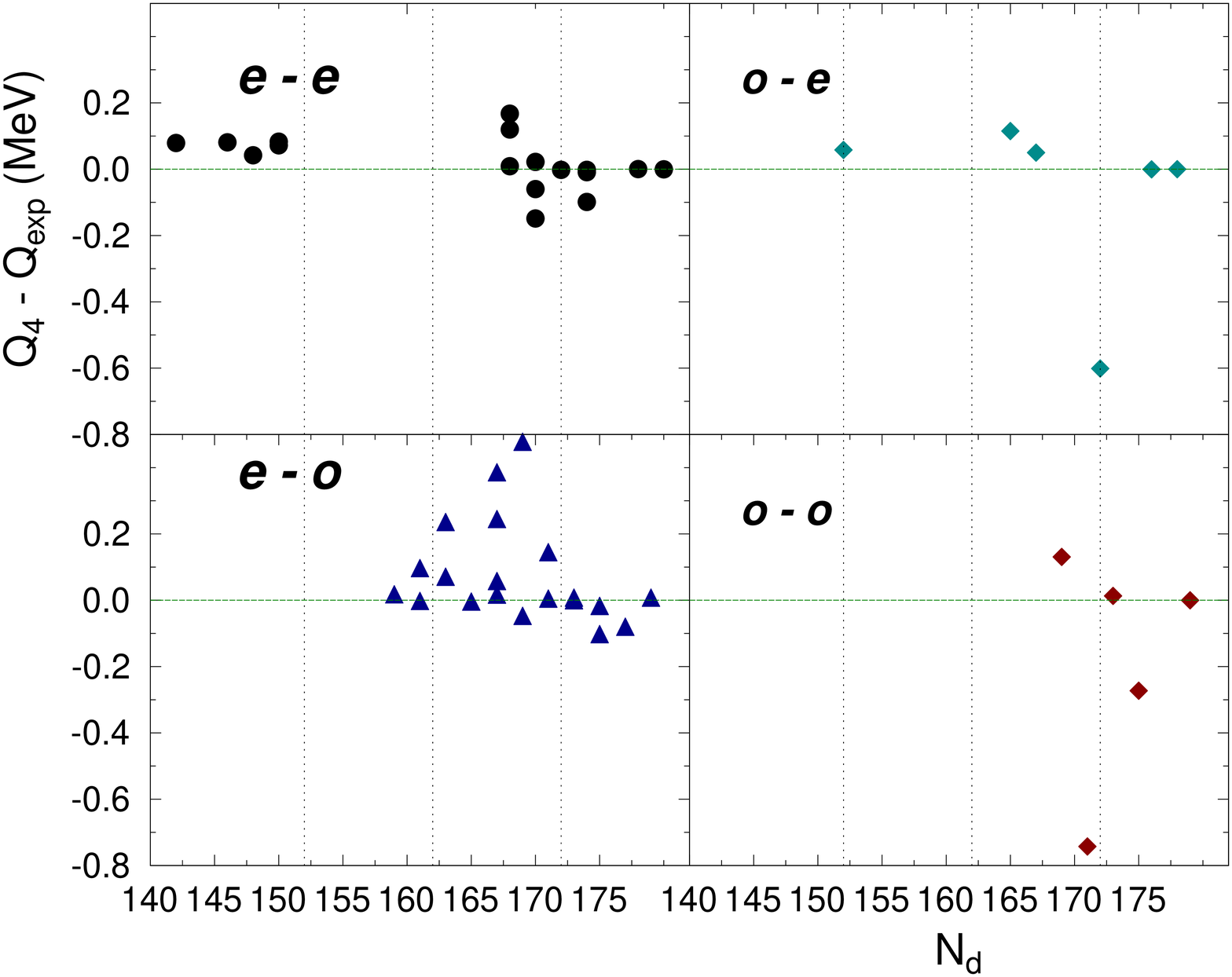} %
\end{center}
\caption{The differences of $Q_4 - Q_{exp}$ in four groups of nuclides
versus the daughter number of neutrons, $N_d$.
\label{difq4}}
\end{figure}

\begin{figure}[ht]
\begin{center}
\includegraphics[width=15cm]{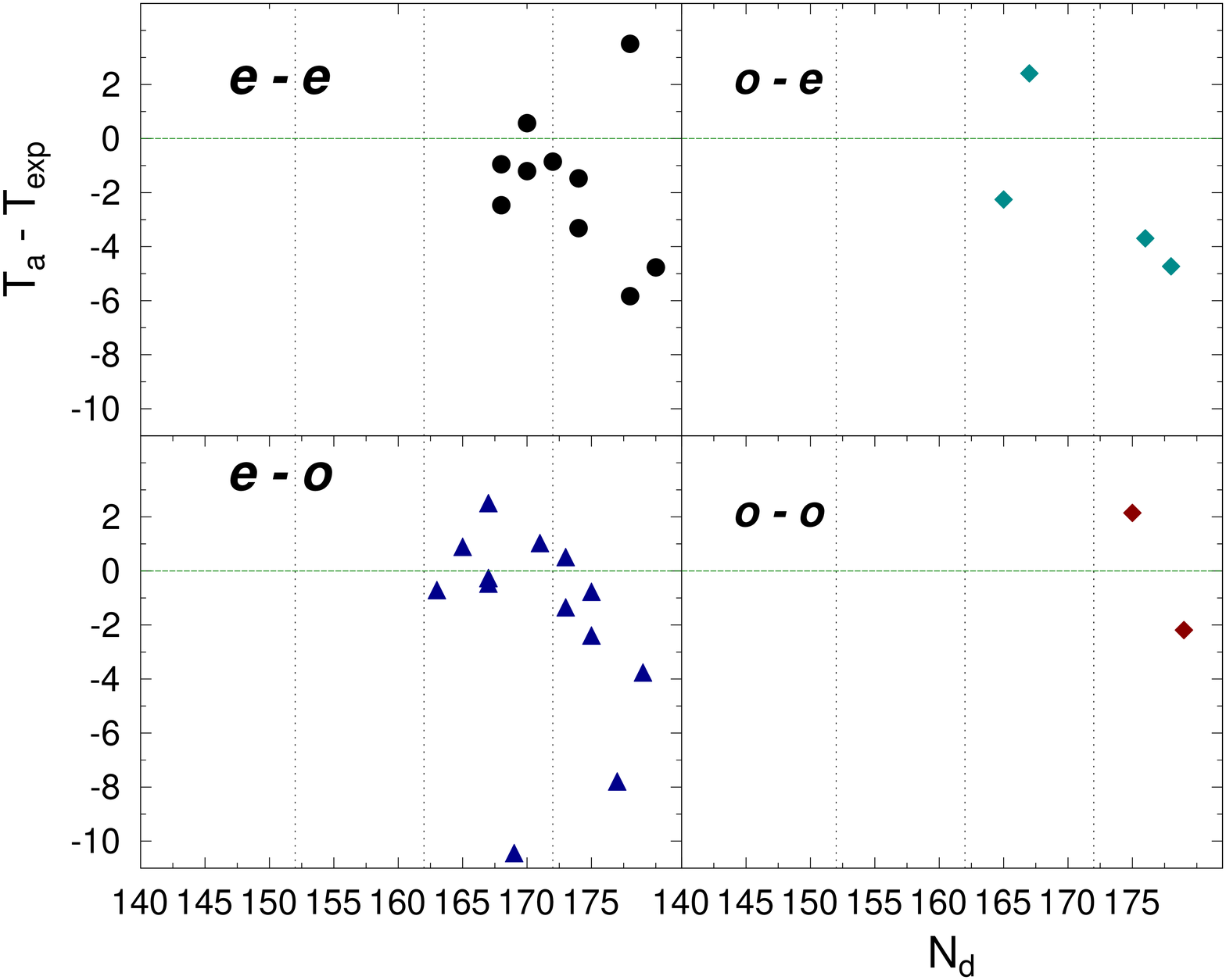} %
\end{center}
\caption{The decimal logarithm of the difference of $T_{semFis} - T_{exp}$
in four groups of nuclides versus the daughter number of neutrons, $N_d$.
\label{dift4}}
\end{figure}

Selected data with the best values are given in the following tables:
\ref{taboee}, \ref{taboeo}, and \ref{tabooeo}.  The corresponding results
are given in the \ref{tabor}.  If we look at the figures~\ref{difq4} and
\ref{dift4} it is very clear that Q-values are very well reproduced, but
generally speaking for the half-lives the errors may rich many orders of
magnitude (9!!!), except few cases: $^{282,284}Cn_{112}$ and
$^{288}Fl_{114}$ (dT under 1) plus $^{284,286,294}Fl_{114}$,
$^{292,294}Og_{118}$ and the evident case of those who have no experimental
data $^{300,302}$120 (dT few units) for e-e nuclei; $^{273,275}Hs_{108}$,
$^{279}Ds_{110}$, $^{281}Cn_{112}$, $^{287,289}Fl_{114}$, $^{293}Lv_{116}$
(dT under 1) plus $^{277}Hs_{108}$, $^{291}Lv_{116}$, $^{295}Og_{118}$,
$^{299,301}$120 for e-o; $^{277}Hs_{108}$, $^{277}Ds_{110}$,
$^{277}Hs_{108}$, $^{277}Ds_{110}$, $^{297,299}$119 (dT few units) for o-e,
and $^{290,300}$119.  The differences $DQ=Q_4 - Q_{exp}$ between Q$_4$, or
Q$_{KTUY05}$ when Q$_4$ is not available are generally speaking quite small,
except for the following cases: $DQ=-4.721, -5.339$~MeV, respectively, for
$^{288}Fl_{114}$ out of 13 e-e nuclides.  For 18 e-o $\alpha$ emitters is
much simpler to give few cases with small $DQ=-0.725, -0.605$ for
$^{275,277}Hs_{108}$ Particularly large $DQ=-4.948$ is observed for
$^{279}Ds_{110}$.  Similarly for 4 odd-even nuclides the only one with small
$DQ=0.283$ is $^{253}Es_{99}$.
For 9 of the $\alpha$ emitters we found AME16 data, which are
given in the table~\ref{bame16}.

In the following part we shall study the behaviour of the optimized values
of Q and T; just one line for every nucleus, despite the fact that we can
miss some excited states in this way. The motivation would be that we are
mainly interested in transitions between the ground state states. 

\section{Cluster Radioactivities}

We give in tables \ref{cee} - \ref{coo} the cluster emission with Q-values
calculated using W4 model, and half-lives with ASAF model.  The Q-values are
plotted in figure~\ref{difq4}, and the differences $T_{semFis} - T_{exp}$
in figure~\ref{dift4}.  The most interesting results are those
obtained for the heaviest nuclides: $^{300,302}$120 with branching ratios
$B_\alpha = -0.10$ and 0.49, respectively, $^{299,301}$120 with $B_\alpha =
-1.49$ and -1.17, $^{297,299}$119 with $B_\alpha = -1.99$ and -3.21, and
$^{300}$119 with $B_\alpha = -3.75$.

\section{Q-values}

Compare experiments with AME16 and W4. Differences $\Delta Q = Q_{th} -
Q_{exp}$, and rms standard deviations, $\sigma_{AME16}$ and $\sigma_{W4}$.

\begin{equation}
\log_{10} T_\alpha(s) = - 57.5 \log_{10} {\cal{R}}_\alpha (cm) + C
\end{equation} 
where $C$ depends on the series, e.g. $C=41$ for the $^{238}$U series.
One has approximately ${\cal{R}}_\alpha = 0.325 E_\alpha^{3/2}$ in which the
kinetic energy of $\alpha$~particles, $E_\alpha$, is expressed in MeV and
the range in air, ${\cal{R}}$, in cm. This relationship is now of historical
interest; the effect of atomic number, $Z$, upon decay rate is obscured.
The one-body theory of $\alpha$-decay can explain it and to a good
approximation produces a formula with an explicit dependence on the $Z$
number.  Nowadays, very often a diagram of $\log T_\alpha$ versus
$ZQ^{-1/2}$ is called Geiger-Nuttal plot.

There are many semiempirical relationships allowing to estimate the
disintegration period if the kinetic energy of the emitted particle
$E_\alpha = QA_d/A$ is known.  $Q$ is the released energy and $A_d, A$ are
the mass numbers of the daughter and parent nuclei.  Alpha-decay half-life
of an even-even emitter can also be easily calculated by using the universal
curves or the analytical superasymmetric (ASAF) model.  Some of these
formulae were only derived for a limited region of the parent proton and
neutron numbers.  Their parameters have been determined by fitting a given
set of experimental data.  Since then, the precision of the measurements was
increased and new $\alpha$-emitters have been discovered.

The description of data in the neighborhood of the magic proton and neutron
numbers, where the errors of the other relationships are large, was improved
by deriving a new formula based on the fission theory of $\alpha$-decay
\cite{p83jpl80}.  A computer program \cite{p98cpc82} allows to change
automatically the fit parameters, every time a better set of experimental
data is available.  There are many alpha emitters, particularly in the
intermediate mass region, for which both the Q-values and the half-lives are
well known \cite{men17cpc,aud17cpc}.  Initially it was used a set of 376
data (123 even-even (e-e), 111 even-odd (e-o), 83 odd-even (o-e), and 59
odd-odd (o-o)) on the most probable (ground state to ground state or favored
transitions) $\alpha$-decays, with a partial decay half-life
\begin{equation}
T_\alpha = (100/b_\alpha)(100/i_p)T_t
\end{equation}
where $b_\alpha$ and $i_p$, expressed in percent, represent the
branching ratio of $\alpha$-decay in competition with all other decay
modes, and the intensity of the strongest $\alpha$-transition,
respectively.

In the region of superheavy nuclei the majority of researchers prefer to use
Viola-Seaborg formula.  Recently for nuclei with $Z=84-110$ and $N=128-160$,
for which both $Q_\alpha ^{exp}$ and $T_{exp}$ experimental values are
available, new optimum parameter values \cite{par05app} have been
determined.  The average hindrance factors for 45 o-e ($Z=85-107$), 55 e-o
($Z=84-110$), and 40 o-o ($Z=85-111$, $N=129-161$) nuclei were determined to
be $C_V^p=0.437$, $C_V^n=0.641$, and $C_V^{pn}=1.024$.  In this way
$T_{exp}$ were reproduced by the Viola-Seaborg formula within a factor of
1.4 foe e-e, 2.3 for o-e, 3.7 for e-o and 4.7 for o-o nuclei, respectively. 

Since 1979 one of us (DNP) considered $\alpha$~decay a superasymmetric
fission process.  Consequently a new semiempirical formula for the alpha
decay halflives \cite{p83jpl80} was a straightforward finding.  The
analytical and numerical superasymmetric fission (ASAF \cite{p103jp84} and
NUSAF) models were used together with fragmentation theory developed by the
Frankfurt School, and with penetrability calculations like for
$\alpha$~decay, to predict cluster (or heavy particle) radioactivity
\cite{ps84sjpn80,enc95}.  The extended calculations, e.g.  \cite{p160adnd91}
have been used to guide the experiments and as a reference for many
theoretical developments.  A series of books and chapters in books, e.g. 
\cite{p140b89,p172b96,p195b96,p193b97,p302bb10} are also available.  A
computer program \cite{p98cpc82} gives us the possibility to improve the
parameters of the ASAF model in agreement with a given set of experimental
data.  The UNIV (universal curve) model was updated in 2011
\cite{p308prc11}.

The interest for $\alpha$D is strongly simulated by the search for heavier
and heavier superheavies (SHs) --- nuclides with $Z>103$, produced by fusion
reactions, who may be identified easily if a chain of $\alpha$D leading to a
known nucleus may be measured.  Recently it was shown that for superheavy
nuclei with atomic numbers $Z>121$ \cite{p309prl11,p315prc12} $\alpha$D may
be stronger than CD or spontaneous fission.

A very interesting result was reported by Y.Z.  Wang et al. \cite{wan15prc},
who compared 18 such formulae in the region of superheavy nuclei.  They
found: ``SemFIS2 formula is the best one to predict the alpha-decay
half-lives ...  In addition, the UNIV2 formula with fewest parameters and
the VSS, SP and NRDX formulas with fewer parameters work well in prediction
on the SHN alpha-decay half-lives ...''

\section{Comparison of results obtained with the new formula, semFIS, UNIV,
and ASAF.}

We present the results obtained using the four models. A global indicator
for a given model could be the weighted mean value

\begin{equation}
\sigma_{newF}^B=\frac{21\sigma_{e-e} + 21\sigma_{e-o} + 13\sigma_{o-e} +   
9\sigma_{o-o}}{64} = 4.6065
\end{equation}

Similarly for the other models
\begin{equation}
\sigma_{ASAF}=\frac{21\sigma_{e-e} + 21\sigma_{e-o} + 13\sigma_{o-e} +   
9\sigma_{o-o}}{64} =4.9477
\end{equation}

\begin{equation}
\sigma_{UNIV}=\frac{21\sigma_{e-e} + 21\sigma_{e-o} + 13\sigma_{o-e} +   
9\sigma_{o-o}}{64} = 4.6476
\end{equation}

\begin{equation}
\sigma_{semFIS}=\frac{21\sigma_{e-e} + 21\sigma_{e-o} + 13\sigma_{o-e} +   
9\sigma_{o-o}}{64} =5.4519
\end{equation}

The rms standard deviations obtained with all models are compared in
table~\ref{rmsall}.

From the results in table ~\ref{rmsall}, we may say that, unexpectedly
semFIS came this time on the last global position.  AKRA is the best,
followed by UNIV and ASAF.  Once again, we may see how important could be
the experimental set of data we are dealing with.  In order to make it very
clear how much the result may depend on the quality of experimental data we
reproduce from a previous publication \cite{p311jpg12}:

\begin{equation}
\sigma_{semFIS534}=\frac{173\sigma_{e-e} + 134\sigma_{e-o} + 123\sigma_{o-e} +   
104\sigma_{o-o}}{534} =0.40803
\end{equation}

\section{Possible Chains of Heaviest SHs}

\begin{figure}[ht]
\begin{center}
\includegraphics[width=15cm]{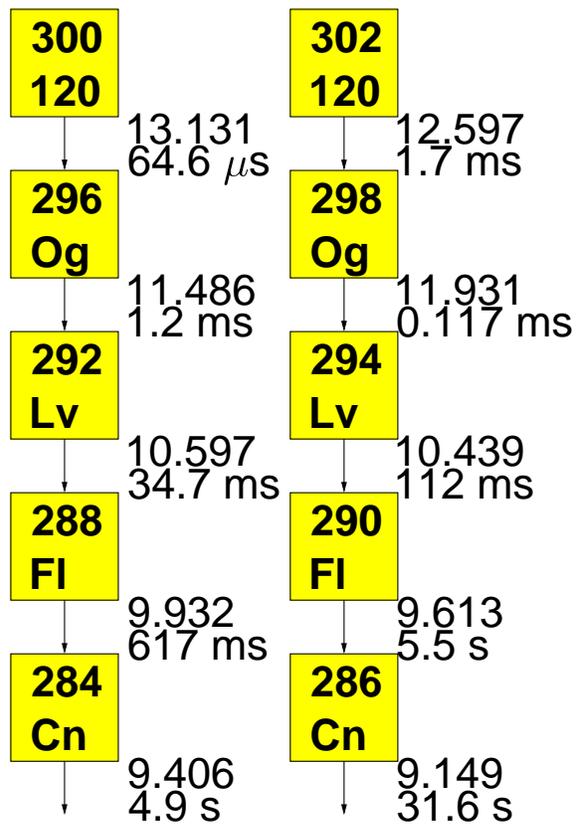} %
\end{center}
\caption{Few possible alpha decay chains of even-even SH emitters.
We give the kinetic energy (MeV) and the half-life of the parent nucleus. 
\label{ale}}
\end{figure}

\begin{figure}[ht]
\begin{center}
\includegraphics[width=15cm]{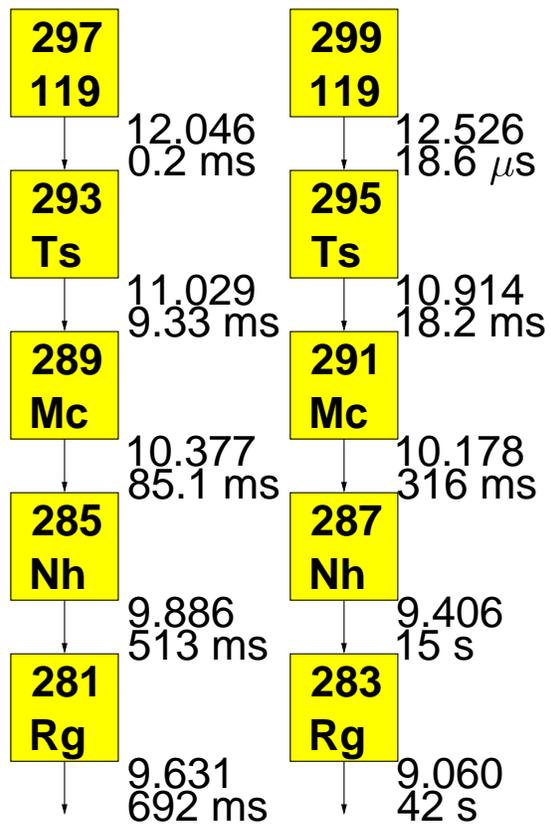} %
\end{center}
\caption{Few possible alpha decay chains of odd-mass SH emitters.
We give the kinetic energy (MeV) and the half-life of the parent nucleus.
\label{alo}}
\end{figure}

We may predict the results shown in the table~\ref{chain} and
figures~\ref{ale},\ref{alo}for $\alpha$D and in table~\ref{clus2} for
cluster radioactivities.

\section{Conclusions}

The accuracy of the new formula was increased after optimization of the five
parameters in the order: a; e; d; c, and b.  The SemFIS formula taking into
account the magic numbers of nucleons, the analytical super-asymmetric
fission model and the universal curves may be used to estimate the alpha
emitter half-lives in the region of superheavy nuclei.  The dependence on
the proton and neutron magic numbers of the semiempirical formula may be
exploited to obtain informations about the values of the magic numbers which
are not well known until now.

We introduced a weighted mean value of the rms standard deviation, allowing
to compare the global properties of a given model.  In this respect for the
set B the order of the four models is the following: semFIS; UNIV; newF, and
ASAF.  

The quality of experimental data was also tested, as one can see by
comparing the three sets (A, B, C).  The set B with large number of emitters
(580) gives the best global result.  It is followed by the set A (454) three
times and the set C (344).
Despite its simplicity in comparison with semFIS the new formula, presented
in this article, behaves quite well, competing with the others well known
relationships discussed in the Ref. ~\cite{wan15prc}. 

We made few predictions concerning possible $\alpha$D decay chains of future
SHs.  For $^{92,94}$Sr cluster radioactivity of $^{300,302}$120 we predict a
branching ratio relative to $\alpha$~decay of -0.10 and 0.49, respectively,
meaning that it is worth trying to detect such kind of decay modes in
competition with $\alpha$~decay.

\begin{acknowledgments} 

We would like to express our gratitude to Prof Dr Sigurd Hofmann who
gave us his publications we used as input data and for his critical
constructive suggestions allowing to improve our manuscript.

This work was supported within the IDEI Programme under Contracts No. 
43/05.10.2011 and 42/05.10.2011 with UEFISCDI, and NUCLEU Programme
PN16420101/2016 Bucharest.  

\end{acknowledgments}


\newpage

\newpage

\begin{table}[ht] 
\caption{Possible Cluster decay modes in competition with $\alpha$ decay 
for the Heaviest SHs.
\label{clus2}}       
\begin{center}
\begin{ruledtabular}
\begin{tabular}{ccccc}
Parent& Emitted    & E$_c$(MEV)& T$_c$ ($\mu s$)& $B_\alpha$ \\  \hline
300 120& $^{92}$Sr & 175.677   &   1.86         & -0.10       \\
302 120& $^{94}$Sr & 175.916   &   5.50         &  0.49       \\
\end{tabular}
\end{ruledtabular}
\end{center}
\end{table} 

\begin{table}[ht] 
\caption{Possible Chains of $\alpha$ decay for the Heaviest SHs.
\label{chain}}       
\begin{center}
\begin{ruledtabular}
\begin{tabular}{ccc}
Parent & E$_\alpha$ (MeV) & T$_\alpha$ \\  \hline
297 119& 12.046&   0.200 ms             \\
293 117& 11.029&   9.330 ms             \\
289 115& 10.377&   85.10 ms \\
285 113&  9.886&   513   ms \\
281 111&  9.631&   692   ms \\
277 109&  9.570&   251   ms \\
273 107&  8.927&    5 s     \\

299 119& 12.526& 18.6 $\mu$s \\
295 117& 10.914& 18.2 ms \\
291 115& 10.178& 316  ms \\
287 113&  9.406& 15   s \\
283 111&  9.060& 42 s \\

300 119& 12.278& 6.46 ms \\
296 117& 11.209& 64.6 ms \\
292 115&  9.912& 5 min 31s \\
288 113&  9.493& 25 min 49s\\
284 111&  8.760& 6days 6h 27 min\\

300 120& 13.131& 64.6 $\mu$s \\
296 118& 11.486&  1.20 ms \\
292 116& 10.597&  34.7 ms \\ 
288 114&  9.932&  617 ms \\  
284 112&  9.406&  4.9 s \\

302 120& 12.597& 1.70 ms \\
298 118& 11.931& 0.117 ms \\
294 116& 10.439& 112 ms \\
290 114&  9.613& 5.5 s \\
286 112&  9.149& 31.6 s \\
\end{tabular}
\end{ruledtabular}
\end{center}
\end{table} 

\begin{table}[ht] 
\caption{Comparison of rms standard deviations. $\sigma$, obtained with
different models.
\label{rmsall}}       
\begin{center}
\begin{ruledtabular}
\begin{tabular}{ccccc}
Parity&AKRA&ASAF&  UNIV & semFIS \\  \hline
e-e&2.989   &3.397&2.952&3.177\\
e-o&8.062   &8.458&8.146&9.453\\
o-e&3.881   &4.056&3.988&4.439 \\
o-o&1.366   &1.663&1.547&2.885 \\
Global&4.606&4.948&4.648&5.4519 \\
\end{tabular}
\end{ruledtabular}
\end{center}
\end{table} 

\begin{table}[ht] 
\caption{Half-lives, $\log_{10} T_\alpha (s)$,  of few nuclides, as given in
the Ref.~\cite{aud12cpc1}.
\label{aud12}}       
\begin{center}
\begin{ruledtabular}
\begin{tabular}{ccc}
A&Z& $\log_{10} T_\alpha (s)$\\  \hline
252&100&4.961\\
269&106&2.681\\
273&108&-0.041\\
277&110&-1.658\\
281&110&1.076\\
281&112&-0.432\\
285&112&1.505\\
285&114&-0.328\\
287&114&-0.284\\
289&114&0.380\\
291&116&-1.553\\
293&116&-1.097\\
253&99&6.248\\
\end{tabular}
\end{ruledtabular}
\end{center}
\end{table} 

\begin{table}[ht] 
\caption{Optimization of coefficients using the set B.
\label{tabcb}}       
\begin{center}
\begin{ruledtabular}
\begin{tabular}{cccccccc}
Group&n&$\sigma$&a&b&c&d&e\\  \hline
e-e & 188 &0.540   &-27.837&-0.94199975&1.5343 &-5.7004 &8.785 \\ 
e-o & 147 &0.678& -28.2245 & -0.8629   &1.53774&-21.145 &53.890 \\
o-e & 131 &0.522&  -26.8005& -1.10783  & 1.5585&14.8525 &-30.523 \\
o-o & 114 &0.840&  -23.6354& -0.891    & 1.404 &-12.4255& 36.9005\\  
\end{tabular}
\end{ruledtabular}
\end{center}
\end{table} 

\begin{table}[ht] 
\caption{Cluster radioactivities of even-even emitters. Q-values obtained
using W4 model, and half-lives with ASAF model.
\label{cee}}  
\begin{center}
\begin{ruledtabular}
\begin{tabular}{ccccccc}
A & Z & A$_e$  & Z$_e$ & Q$_c$ (MeV) & $\log_{10} T_c(s)$ & B$_a=T_\alpha - T_c$  
\\  \hline
252 &100&48&20[Ca]&145.85&23.63&-20.88 \\
278 &108&72&28[Ni]&216.64&16.76&-15.51 \\
282 &110&74&28[Ni]&223.06&15.21&-12.89 \\
282 &112&74&30[Zn]&245.52&9.29&-10.24 \\
284 &112&76&30[Zn]&245.30&8.91&-8.23 \\
284 &114&78&32[Ge]&264.41&6.71&-9.18 \\
286 &114&80&32[Ge]&264.23&6.18&-7.22 \\
288 &114&80&32[Ge]&264.72&5.12&-5.33 \\
290 &114&82&32[Ge]&263.89&5.30&-4.56 \\
294 &114&82&32[Ge]&258.17&10.81&-7.31 \\
292 &116&84&34[Se]&284.64&0.55&-2.01 \\
294 &118&86&36[Kr]&303.81&-2.45&-0.87 \\
300 &120&92&38[Sr]&321.36&-5.73&-0.10 \\
302 &120&94&38[Sr]&320.04&-5.26&0.49 \\
\end{tabular}
\end{ruledtabular}
\end{center}
\end{table}

\begin{table}[ht] 
\caption{Cluster radioactivities of even-odd emitters. Q-values obtained
using W4 model, and half-lives with ASAF model.
\label{ceo}}  
\begin{center}
\begin{ruledtabular}
\begin{tabular}{ccccccc}
A &Z & A$_e$  & Z$_e$ &Q$_c$ (MeV) &$\log_{10} T_c(s)$ & B$_a=T_\alpha - T_c$  
\\  \hline
265 &104&55&22[Ti]&165.27&26.71&-22.00 \\
267 &104&61&24[Cr]&175.93&28.83&-24.30 \\
269 &106&64&26[Fe]&195.84&24.94&-22.56 \\
271 &106&65&26[Fe]&195.65&24.86&-23.00 \\
273 &108&68&28[Ni]&216.27&20.25&-20.30 \\
275 &108&70&28[Ni]&216.20&19.97&-18.98 \\
277 &108&71&28[Ni]&216.04&19.76&-17.35 \\
279 &110&71&28[Ni]&225.09&15.77&-15.84 \\
281 &110&72&28[Ni]&223.55&17.05&-14.59 \\
279 &110&71&28[Ni]&225.09&15.77&-15.84 \\
281 &112&74&30[Zn]&245.18&12.15&-12.59 \\
283 &112&76&30[Zn]&244.79&12.00&-10.47 \\
285 &112&77&30[Zn]&244.08&12.26&-8.81 \\
287 &114&80&32[Ge]&264.49&8.04&-6.56 \\
289 &114&81&32[Ge]&263.78&8.22&-5.94 \\
291 &116&84&34[Se]&284.42&3.58&-5.01 \\
293 &116&85&34[Se]&283.13&4.34&-5.15 \\
295 &118&87&36[Kr]&303.06&0.50&-2.73 \\
299 &120&91&38[Sr]&321.48&-2.70&-1.49 \\
301 &120&93&38[Sr]&320.58&-3.86&-1.17 \\
\end{tabular}
\end{ruledtabular}
\end{center}
\end{table}

\begin{table}[ht] 
\caption{Cluster radioactivities of odd-even and odd-odd emitters. Q-values
obtained using W4 model, and half-lives with ASAF model.
\label{coo}}  
\begin{center}
\begin{ruledtabular}
\begin{tabular}{ccccccc}
A &Z & A$_e$  & Z$_e$ &Q$_c$ (MeV) &$\log_{10} T_c(s)$ & B$_a=T_\alpha - T_c$  
\\  \hline
253 & 99&46&18[Ar]&129.54&25.87&-19.44 \\
277 &108&71&18[Ar]&216.04&19.76&-17.35 \\
277 &108&71&28[Ni]&225.98&15.24&-17.49 \\
287 &113&79&31[Ga]&254.02&8.97&  -7.80 \\
297 &119&89&37[Rb]&311.65&-1.71&-1.99 \\
299 &119&91&37[Rb]&310.63&-1.52&-3.21 \\
278 &107&73&28[Ni]&211.19&22.73&-15.79 \\
282 &109&71&27[Co]&208.28&25.44&-17.78 \\
286 &111&78&29[Cu]&230.34&18.88&-12.76 \\
290 &113&81&31[Ga]&251.27&13.45&-9.30 \\
300 &119&92&37[Rb]&309.74&1.56&-3.75 \\
\end{tabular}
\end{ruledtabular}
\end{center}
\end{table}

\begin{table}[ht] 
\caption{Best results obtained with AME16 for $\alpha$~emitters. 
\label{bame16}}  
\begin{center}
\begin{ruledtabular}
\begin{tabular}{ccccc}
A &Z &Q$_{exp}$ (MeV) & Q$_2$ (MeV) & DQ= Q$_2 -$ Q$_{exp}$ (MeV) \\  \hline
282&112&10.10&10.10&0.899E-02 \\
288&114&10.10&10.10&-0.201E-02 \\
290&116&11.00&11.0&-0.801E-02 \\
292&116&10.80&10.80&-0.200E-02 \\
294&118&11.80&11.80&-0.900E-02 \\
281&112&10.40&10.40&0.900E-02 \\
285&112&~9.32&~9.32&-0.300E-02 \\ 
287&114&10.20&10.20&-0.500E-02 \\
291&116&10.90&10.90&{\bf -0.954E-06} \\
\end{tabular}
\end{ruledtabular}
\end{center}
\end{table}

\begin{table}[ht] 
\caption{Best results obtained with W4 for $\alpha$~emitters. 
\label{bw4}}  
\begin{center}
\begin{ruledtabular}
\begin{tabular}{cccccccc}
A &Z & Q$_{exp}$ (MeV) & Q$_4$ (MeV) & DQ= Q$_4 -$ Q$_{exp}$ &
$ log_{10}T_{semFIS}$ & $ log_{10}T_{ASAF}$
& DT=T$_{semFIS} -  T_{ASAF}$\\  \hline
282&112&10.10&10.10&0.899E-02&-1.35&-0.952&0.236\\
288&114&10.10&10.10&-0.201E-02&-0.702&-0.211&0.155\\
290&116&11.10&11.10&-0.801E-02&-2.47&-2.02&3.27\\
292&116&10.80&10.80&-0.200E-02&-1.97&-1.46&0.319\\
294&118&11.80&11.80&-0.900E-02&-3.84&-3.31&0.319\\
281&112&10.40&10.50&0.900E-02 &-1.63&-0.442&0.479\\
285&112& 9.32& 9.32&-0.300E-02&1.49  &3.45&0.229\\
287&114&10.20&10.20&-0.500E-02&-0.299&1.47&0.208\\
289&114& 9.97& 9.97&-0.801E-02&0.208&0.228&0.228\\
291&116&10.90&10.90&{\bf -0.954E-06}&-1.62&-1.43&0.592\\
\end{tabular}
\end{ruledtabular}
\end{center}
\end{table}

{\small
\begin{table}[ht] 
\caption{Selected best input data with half-lives in decimal logarithm of
the $T_{1/2}$ expressed in seconds, $lgT=\log_{10}T_{1/2}(s)$.  Even-even
$\alpha$~emitters.
\label{taboee}}       
\begin{center}
\begin{ruledtabular}
\begin{tabular}{cccccccccccc}
Crt. nb.&A&Z&N$_d$&T$_{med}$&Q (MeV)& T$_{min}$&T$_{max}$& $lgT_{med}$& 
$lgT_{min}$&$lgT_{max}$& $\Delta T$ \\  \hline
1  &240 & 96 &142&$2.3328\times 10^6$&6.398& & &6.368& & & \\
2  &246 & 98 &146&$1.2852\times 10^5$ &6.862& & &5.109& & & \\
3  &248 & 98 &148&$2.87712\times 10^7$&6.361& & &7.459& & & \\
4  &250 & 98 &150&$9.839232\times 10^9$&6.128& & &9.993& & & \\
5  &252 &100 &150&$9.0\times 10^4$&7.153& & &4.954& & & \\
6  &278 &108 &168&690.0&8.8&380.0&3990.0&2.839&2.580&3.601&1.021\\
7  &282 &110 &170&67.0&8.96&37.0&387.0&1.826&1.568&2.588&1.020\\
8  &284 &112&170&0.118&9.605&0.101&0.142&-0.928&-0.996&0.148\\
9  &286 &112&172&640.&8.793&340.&3740.&2.806&2.531&3.573&1.041\\
10 &284 &114&168&0.002&10.76&0.0013&0.0047&-2.699&-2.886&-2.328&0.558\\
11 &286 &114&170&0.166&10.65&     &     &-0.780&   &         &     \\
12 &288 &114&172&0.644&14.795&0.547&0.782&-0.191&-0.262&-0.107&0.155\\             
13 &290 &114&174&21.0&9.846&11.0&122.0&1.322&1.041&2.086&1.045\\             
14 &294 &114&178&0.002&10.760&     &     &-2.699&   &         &     \\             
15 &290 &116&172&0.0083&11.005&0.0064&11.9&-3.495&-3.658&-3.237&0.421\\             
16 &292 &116&174&0.0128&10.776&0.0095&0.0198&-1.893&-2.022&-1.703&0.319\\             
17 &294 &118&174&0.0022&11.835&0.0012&12.7&-2.658&-2.921&1.104  \\
18 &300 &120&178&0.0   &13.308&     &     &-5.83&   &    &     \\
19 &302 &120&180&0.0   &12.766&     &     &-4.77&   &    &     \\
\end{tabular}
\end{ruledtabular}
\end{center}
\end{table}} 

{\small
\begin{table}[ht] 
\caption{Selected best input data with half-lives in decimal logarithm of
the $T_{1/2}$ expressed in seconds, $lgT=\log_{10}T_{1/2}(s)$.  Even-odd
$\alpha$~emitters.
\label{taboeo}}       
\begin{center}
\begin{ruledtabular}
\begin{tabular}{cccccccccccc}
Crt. nb.&A&Z&N$_d$&T$_{med}$&Q (MeV)& T$_{min}$&T$_{max}$& $lgT_{med}$& 
$lgT_{min}$&$lgT_{max}$& $\Delta T$ \\  \hline
1  &265 &104 &159&61.   & 6.398&39.00&145.0&1.785&1.591&2.161&0.570 \\
2  &267 &104 &161&4608. & 6.862&2808.&12996.&3.664&3.448&4.114&0.665 \\
3  &269 &106 &161&185.0 & 6.361&117.0&439.0 &2.267&2.068&2.642&0.574 \\
4  &273 &108 &163&0.765&  7.153& 0.  &  0.  &-0.116&1.806&2.283&0.477 \\
5  &275 &108 &165&0.201& 10.175& 0.& 0.&-0.873&-0.396&0.477&0. \\
6  &277 &108 &167&0.0031& 9.605&1.70&18.0&-2.509&0.230&1.255&1.025\\
7  &279 &110 &167&0.290& 10.795&0.243&0.359&-0.538&-0.614&-0.445&0.169\\
8  &281 &110&169&13.0&   10.365&10.3&17.50&1.114&1.013&1.243&0.230\\
9  &279 &110&167&0.290&  14.795&0.  &0.  &-0.538&1.013&1.243&0.230\\
10 &281 &112&167&0.128&  11.005&0.085&0.256&-0.893&-1.071&-0.592&0.479\\
11 &283 &112&169&308.&   16.115&219. &520. &2.489&2.340&2.716& 0.376\\
12 &285 &112&171&28.9&   11.835&23.0&39.00&1.461&1.362&1.591&0.229\\             
13 &287 &114&171&0.540&  10.025&0.440&0.710&-0.268&-0.357&-0.149&0.208\\             
14 &289 &114&173&1.870 &  7.805& 1.49&2.52&0.272&0.173&0.401&0.228\\             
15 &291 &116&173&0.0180&  7.885&0.0   &0.0   &-1.745&-1.959&-1.367&0.592\\             
16 &293 &116&175&0.0570&  8.645&0.0390&0.103&-1.244&-1.409&-0.987&0.422\\             
17 &295 &118&175&0.261&   8.895&0.0 & 0. &-0.583&-1.409&-0.987&0.422\\
18 &299 &120&177&3.7 &13.14&0.   &0.      &0.568&-1.009&0.020  &1.029\\
19 &301 &120&179&0.0 &12.939&0.  &0.      &-3.86&      &0.     &0.  \\
\end{tabular}
\end{ruledtabular}
\end{center}
\end{table}} 

{\small
\begin{table}[ht] 
\caption{Selected best input data with half-lives in decimal logarithm of
the $T_{1/2}$ expressed in seconds, $lgT=\log_{10}T_{1/2}(s)$.  Odd-even and
odd-odd  $\alpha$~emitters. For $^{297,299}$119 we have used the KTUY mass
model \cite{kou05ptp}.
\label{tabooeo}}       
\begin{center}
\begin{ruledtabular}
\begin{tabular}{cccccccccccc}
Crt. nb.&A&Z&N$_d$&T$_{med}$&Q (MeV)& T$_{min}$&T$_{max}$& $lgT_{med}$& 
$lgT_{min}$&$lgT_{max}$& $\Delta T$ \\  \hline
1  &253 & 99 &152&1771200.& 6.398&0. &0. &6.248&0. &0. &0. \\
2  &277 &108 &167&0.0031&   6.862&1.7&18.&-2.509&0.230&1.255&1.02 \\
3  &277 &110 &165&0.0041&   6.361&0. &0. &-2.387&0. &0. &0.0 \\
4  &287 &113 &172&14.00 &   6.128&0. &0. &1.146&0. &0. &0. \\
5  &297 &119 &176&12.210&  11.285&0. &0. &-3.70&0. &0. &0.\\
6  &299 &119 &178&12.696&  11.475&0. &0. &-4.73&0. &0. &0.\\
7  &278 &107 &169&690.0 &7.900&380.&3990.&2.839&2.580&3.601&0.\\
8  &282 &109 &171&67.0  &8.83 &0.  &0.   &1.826&0.   &0.   &0.\\
9  &286 &111 &173&642.  &8.67 &0.  &0.   &2.808&0.   &0.   &0.\\
10 &290 &113 &175&2.0   &9.71 &0.0 &0.   &0.301&0.   &0.   &0.\\
11 &300 &119 &179&0.    &12.444&0. &0.   &-2.19&0.   &0.   &0.\\
\end{tabular}
\end{ruledtabular}
\end{center}
\end{table}} 

\begin{table}[ht] 
\caption{Best results obtained with selected $\alpha$~emitters. Q-values
given by W4 model. 
\label{tabor}}  
\begin{center}
\begin{ruledtabular}
\begin{tabular}{cccc}
n &parity&$\sigma_Q$ & $\sigma_{T-ASAF}$ \\  \hline
16&e-e&0.489&0.448 \\
19&e-o&0.171&1.280 \\
 4&o-e&0.356&0.986 \\
 4&o-o&0.463&2.920 \\
\end{tabular}
\end{ruledtabular}
\end{center}
\end{table}

\end{document}